\documentclass[runningheads]{llncs}
\usepackage{graphicx}
\usepackage{subfigure}
\begin{document}
%
\title{6GCVAE: Gated Convolutional Variational Autoencoder for IPv6 Target Generation}
\titlerunning{Gated Convolutional Variational Autoencoder for IPv6 Target Generation}
%

\author{}
\author{Tianyu Cui\inst{1,2} \and
Gaopeng Gou\inst{1,2,}\thanks{Gaopeng Gou is the corresponding author, email address:  gougaopeng@iie.ac.cn.} \and
Gang Xiong\inst{1,2}}

\authorrunning{T. Cui et al.}
%

\institute{}
\institute{Institute of Information Engineering, Chinese Academy of Sciences \and
School of Cyber Security, University of Chinese Academy of Sciences\\
Beijing, China\\
\email{ \{cuitianyu, gougaopeng, xionggang\}@iie.ac.cn}}

%
\maketitle              
\begin{abstract}
IPv6 scanning has always been a challenge for researchers in the field of network measurement.  Due to the considerable IPv6 address space, while recent network speed and computational power have been improved, using a brute-force approach to probe the entire network space of IPv6 is almost impossible. Systems are required an algorithmic approach to generate more possible active target candidate sets to probe. In this paper, we first try to use deep learning to design such IPv6 target generation algorithms. The model effectively learns the address structure by stacking the gated convolutional layer to construct Variational Autoencoder (VAE). We also introduce two address classification methods to improve the model effect of the target generation. Experiments indicate that our approach 6GCVAE outperformed the conventional VAE models and the state of the art target generation algorithm in two active address datasets.


\end{abstract}

\section{Introduction}
In the network measurement task, in order to discover the active hosts in the network and judge their active state, the researchers usually use the network scanning method to actively detect all the hosts
existing in the network space. Systems confirm that the host is active by sending the request packets and waiting until receiving the response packets from the host. However, IPv6 \cite{deering1998internet} contains a considerable address space. The current scanner \cite{durumeric2013zmap} cannot complete the entire IPv6 network space scanning.

The state of the art approach to solving this problem is using IPv6 target generation technology \cite{ullrich2015reconnaissance,foremski2016entropy,murdock2017target}. The technology requires a set of active IPv6 seed addresses as the input and learns the structure of the seed addresses to generate possible active IPv6 target candidate sets. Due to the semantics of the IPv6 address is opaque, it is difficult to infer the IPv6 address structure of a real host or perform effective analysis of the addressing schemes.

The representative algorithms of IPv6 target generation technology include Entropy/IP \cite{foremski2016entropy} which trained the Bayesian network to generate active candidate sets. However, the approach requires assuming that address segments exist dependency. The confirmed model determined by experience and assumption may be influenced in various datasets, thus leading to quite different effects \cite{foremski2016entropy}. In addition, because of the characteristics of such algorithms, they will consume a long time under a large dataset.

Deep neural network architectures are used for the batch processing of big data tasks. Models are able to automatically adapt to seed datasets by training, thus usually performing well in a variety of large datasets. Variational Autoencoder (VAE) \cite{kingma2013auto} is a typical generative model in deep neural networks. The model samples the latent vector and finally reconstructs the text or image that is similar to the original. 
The encoding idea may contribute to deeply mine the potential relationship between addresses and active hosts. The gated convolutional network was proposed by Dauphin et al. \cite{dauphin2017language} The convolution and gating mechanism of the model effectively learn the text structure while understanding the relevance of the text, which can help models learn the key features of IPv6 addresses.

In this paper, we use a deep neural network architecture for the first time to accomplish the IPv6 target generation task. Our contribution can be summarized as follows:
\vspace{-0.2cm}
\begin{itemize} 
\item We first propose using deep learning architecture to achieve IPv6 target generation. Our work achieves a new model 6GCVAE that stacks the gated convolutional layer to construct VAE model.
\item We use two methods of seed classification, which contributes to explore the IPv6 addressing schemes to effectively improve the effect of the model.
\item Our model demonstrates better results on both two datasets than conventional VAE models  (FNN VAE, Convolutional VAE, RNN VAE, LSTM VAE, and GRU VAE) and the state of the art target generation technology Entropy/IP.
\end{itemize}

%
\vspace{-0.2cm}
The organizational structure for the rest of the paper is as follows. Section~2 introduces the related work of IPv6 target generation. Section 3 introduces the background and considerations of this task. 6GCVAE architecture and seed classification methods are shown in  Section 4. Section 5 evaluates our work and Section 6 summarizes the paper.
\vspace{-0.3cm}

\section{Related Work}
\vspace{-0.2cm}
In previous work, researchers have found there are certain patterns in active IPv6 address sets. Planka and Berger \cite{plonka2015temporal} first explored the potential patterns of IPv6 active addresses in time and space. They used Multi-Resolution Aggregate plots to quantify the correlation of each portion of an address to grouping addresses together into dense address space regions.  Czyz et al. \cite{czyz2016don} found 80\% of the routes and 22\% of the server addresses have only non-zero addresses in the lowest 16 bits of the address. Gasser et al. \cite{gasser2018clusters} used entropy clustering to classify the hitlist into different addressing schemes. We adopt their methods by performing seed classification to help neural networks improve model performance.

Ullrich et al. \cite{ullrich2015reconnaissance} used a recursive algorithm for the first attempt to address generation. They iteratively searched for the largest match between each bit of the address and the current address range until the undetermined bits were left, which is used to generate a range of addresses to be scanned. Murdock et al. \cite{murdock2017target} introduced 6Gen, which generates the densest address range cluster by combining the closest Hamming distance addresses in each iteration. Foremski et al. \cite{foremski2016entropy} used Entropy/IP for efficient address generation. The algorithm models the entropy of address bits in the seed set and divides the bits into segments according to the entropy values. Then they used a Bayesian network to model the statistical dependence between the values of different segments. This learned statistical model can then generate target addresses for scanning. Different from these work, we use the neural network to construct the generated model and mainly compare it with Entropy/IP.

Researchers have extensively studied the VAE models in many fields, including text generation \cite{semeniuta2017hybrid} and image generation \cite{pu2016variational}. Recently, gated convolutional networks have made outstanding progress on many Natural Language Processing (NLP) tasks due to their parallel computing advantages. Dauphin et al. \cite{dauphin2017language} first proposed the model and called its key modules Gated Linear Units (GLU). Their approach achieves state-of-the-art performance on the WikiText-103 benchmark. Gehring et al. \cite{gehring2017convolutional} simplified the gradient propagation using GLU and made a breakthrough on the WMT'14 English-German and WMT'14 English-French translation. To the best of our knowledge, we are using a gated convolutional network for the first time to construct a VAE model and to overcome the challenge of the IPv6 target generation task.
\vspace{-0.3cm}

\section{IPv6 Target Generation}
In this section, we provide a brief description of IPv6 addressing background and our consideration of target generation tasks. We refer the reader to RFC~2460~\cite{deering1998internet} for a detailed description of the protocol.

\vspace{-0.2cm}

\begin{figure}[t]
\setlength{\belowcaptionskip}{-0.5cm}
\includegraphics[width=\textwidth]{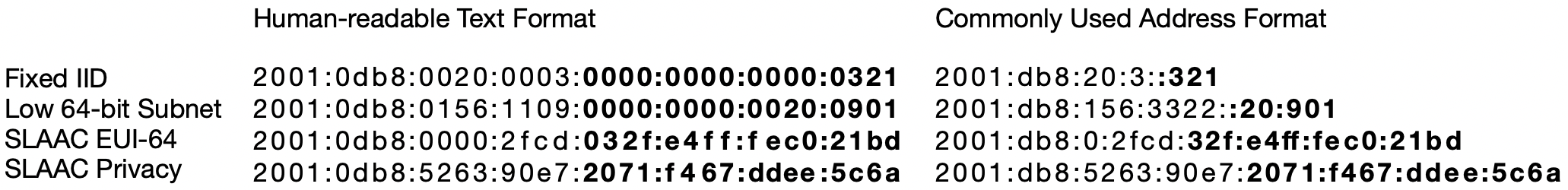}
\caption{Sample IPv6 addresses in presentation format with the low 64 bits shown bold.
} \label{fig1}
\end{figure}

\subsection{IPv6 Addressing Background}
An IPv6 address consists of a global network identifier (network prefix, e.g. /32 prefix), subnet prefix, and an interface identifier (IID) \cite{carpenter2014significance}. It is composed of 128-bit binary digits, which are usually represented in human-readable text format, using 8 groups of 4 hexadecimal digits and separating them by colons, as shown in Figure \ref{fig1}. Each of the hexadecimal digits is called a nybble. Since IPv6 addresses usually use "::" to replace groups of consecutive zero values and omit the first zero value in each group, a commonly used address format representation for IPv6 is also shown in Figure \ref{fig1}.


There are many IPv6 addressing schemes and network operators are reminded to treat interface identifiers as semantically opaque \cite{carpenter2014significance}. Administrators have the option to use various standards to customize the address types. In addition, some IPv6 addresses have SLAAC \cite{narten2007ipv6} address format that the 64-bit IID usually embeds the MAC address according to the EUI-64 standard \cite{narten2007ipv6} or is set completely pseudo-random \cite{narten2001rfc3041}. Consider the sample addresses in Figure \ref{fig1}. In increasing order of complexity, these addresses appear to be:
\vspace{-0.1cm}
\begin{itemize} 
\item an address with fixed IID value (::321).
\item an address with a structured value in the low 64 bits (perhaps a subnet distinguished by ::20).
\item  a SLAAC address with EUI-64 Ethernet-MAC-based IID (ff:fe flag).
\item  a SLAAC privacy address with a pseudorandom IID.
\vspace{-0.1cm}
\end{itemize}

\subsection{Considerations}
\vspace{-0.1cm}

Due to the semantic opacity of IPv6 addresses and the hybridization of multiple addressing schemes, the deep learning model may have difficulty in effectively training when learning the address structure. An address of the SLAAC format also has a highly randomized address structure, which is bound to pose a challenge to the generation task. However, to ensure that each addressing scheme can be included in the generation set, the selected seed set must contain all address patterns. Therefore, the target generation task requires the model to be able to effectively extract the underlying semantic information of IPv6 addresses. In addition, since the mixture of multiple structures, certain classification work on the seed set will alleviate the pressure on the model. 
\vspace{-0.4cm}




\begin{figure}[t]
\setlength{\abovecaptionskip}{0.1cm}
\setlength{\belowcaptionskip}{-0.2cm}
\includegraphics[width=\textwidth]{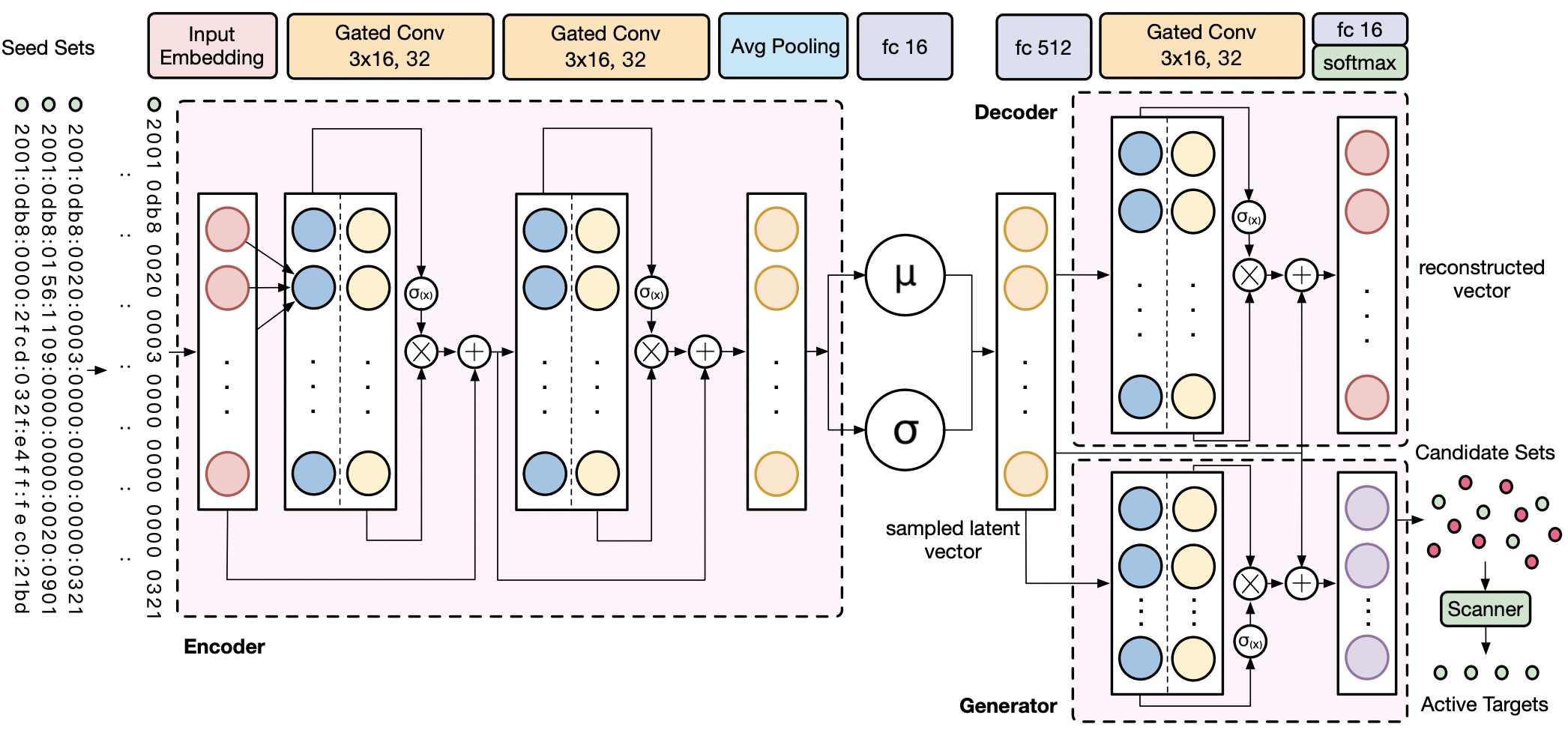}
\caption{The overall architecture of 6GCVAE. The model requires seed sets removed the colon as input and learns the address structure distribution through the encoder. The decoder reconstructs the latent vector after sampling. After training, the generator produces considerable candidates sets waited for probing by a scanner, which can finally discover the active targets.
} \label{fig2}
\vspace{-0.2cm}
\end{figure}

\section{Approach}
\vspace{-0.3cm}
In this section, we will introduce our approach and two seed classification methods for IPv6 target generation.

6GCVAE relies on stacked gated convolutional layers to form a Variational Autoencoder. The detailed model architecture is shown in Figure \ref{fig2}. We remove the colon in each address and leave the 32-bit hexadecimal as a sample input (e.g., 20010db8002000030000000000000301). Since each nybble may be one of {\bfseries 0-f} characters, the alphabet size is 16 and we can arrive at a final input representation with a dimension of 32$\times$16 after input embedding. 

Our training model expects the generated address to be constantly approaching the input address to produce a new possible active target. To achieve the goal, the model is required to learn the distribution of the input by an encoder, sample latent vector and reconstruct the new generation by a decoder.

\begin{figure}[t]
\vspace{-0.3cm}
\begin{center}
\setlength{\abovecaptionskip}{-0.1cm}
\setlength{\belowcaptionskip}{-0.9cm}
\includegraphics[width=0.8\textwidth]{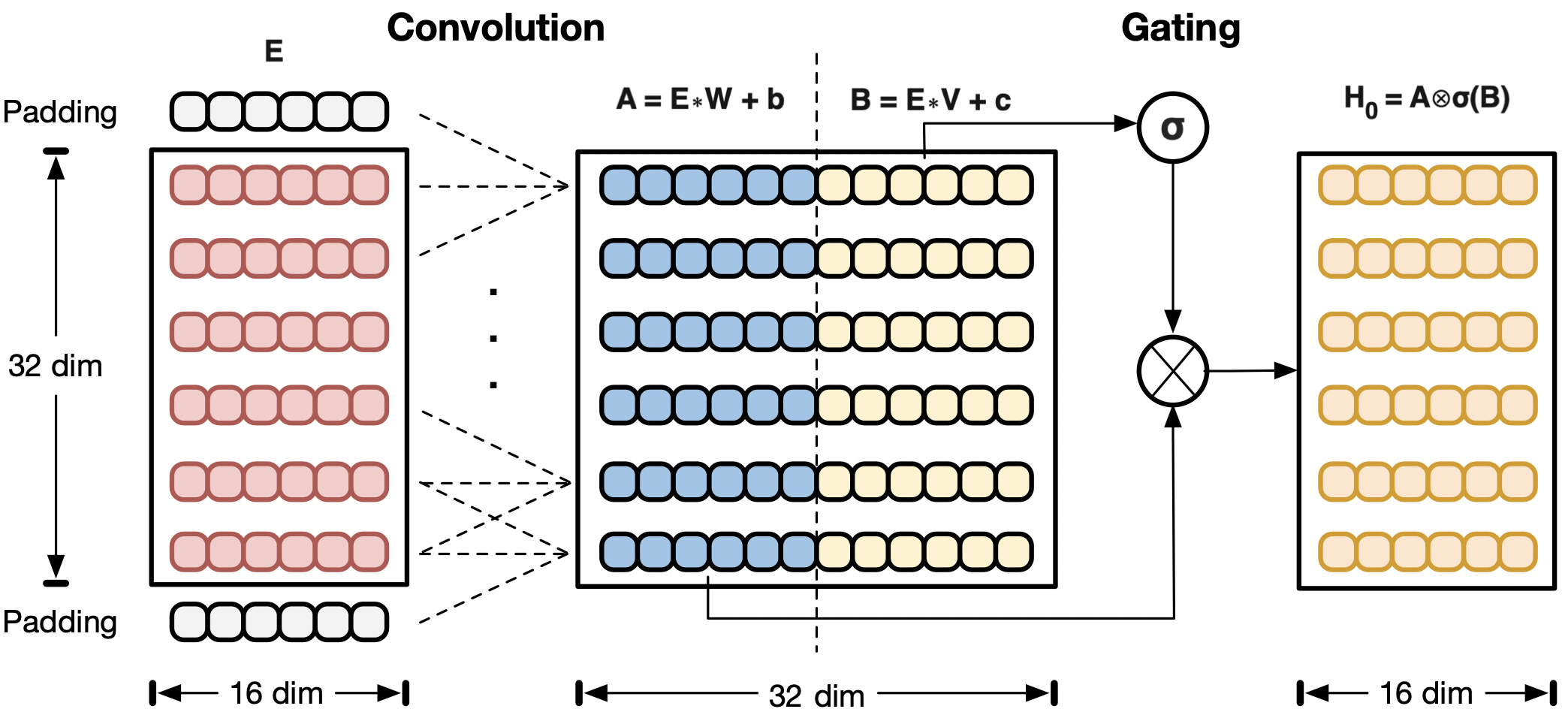}
\caption{Structure of the gated convolutional layer for IPv6 target generation. After convolution, the output of the vector $A$ is controlled by the sigmoid value of vector $B$, which is used as an output gate to select the address vector.
} \label{fig3}
\end{center}
\end{figure}

\vspace{-0.3cm}
\subsection{Gated Convolution Layer}
\vspace{-0.2cm}
The gated convolutional network enables to complete sequence tasks by adding a gating mechanism to the convolution. The structure is shown in Figure \ref{fig3}.

We define the input embedding as $E = [D_0,...,D_i,...,D_{31}]$, where $D_i$ represents the vector representation of the $i$-th nybble of the address.  We use 32 $3\times16$ convolution kernels to convolve the input $E$ to obtain a $32\times32$-dimensional output vector which is divided equally to vector $A$ and vector $B$. Finally, we take the sigmoid function as the gate for the vector $B$ in the second half to control the output of the vector  $A$. The approach to compute the hidden layers $H_i$ can be summarized as

\begin{equation}
H_i = A\otimes\sigma B
\end{equation}

where $\sigma$ is the sigmoid function and $\otimes$ is the element-wise product between matrices.

\vspace{-0.5cm}
\subsubsection{Why Gated Convolution Layer}Using the gating method can effectively help us monitor the importance of each nybble of an IPv6 address. The convolution method also improves the sensitivity of the model to the relationship between each nybble of the address. This allows our model to be able to focus on address importance flags (e.g., the 23rd-26th nybbles of the EUI-64 address are always {\bfseries fffe} ) while discovering potential relationships between address nybbles (e.g., the fixed IID address typically has a contiguous 0).

\vspace{-0.2cm}
\subsection{Variational Autoencoder}
In VAE models,  a deterministic internal representation $z$ (provided by the encoder) of an input $x$ is usually replaced with a posterior distribution $q(z|x)$. Inputs are then reconstructed by sampling $z$ from this posterior and passing them through a decoder. After training, the model will mass-produce text or images by a generator. In this section, we will introduce the encoder, decoder, and generator structure in our approach.
\vspace{-0.5cm}
\subsubsection{Encoder} In our model, we use two gated convolutional layers and an average pooling layer stack as the encoder for the model. In order to maintain the memory of the original input, we used a residual connection between each gated convolutional layer.

According to the principle of VAE, we use two fully connected layers to train the mean $\mu$ and the log variance $log\ \sigma^2$ to learn the distribution of the input x.
\vspace{-0.5cm}
\subsubsection{Decoder} To ensure that we can sample from any point of the latent space and still generate valid and diverse outputs, the posterior $q(z|x)$ is regularized with its KL divergence from a prior distribution $p(z)$. The prior is typically chosen to be a Gaussian with zero mean and unit variance. Then the latent vector $z$ can be computed as

\begin{equation}
z =\mu + \epsilon * \sigma
\end{equation}

where $\epsilon$ is sampled from the prior.

The decoder consists of a gated convolutional layer, fully connected layers, and a softmax activation.  After sampling the latent vector $z$. We use the fully connected layer and adjust it to $32\times16$ dimensions as the input to the gated convolutional layer. Finally, the reconstructed address vector can be obtained through the fully connected layer and softmax activation function.

Our model loss consists of two parts, including the cross-entropy loss $J_{xent}$ and the KL divergence $KL(q(z|x)||p(z))$. The cross-entropy loss expects the smallest reconstruction error between the reconstructed vector y and the input seed x. The KL divergence constraint model samples from the standard normal distribution:

\begin{equation}
J_{xent} = -(x\cdot log(y)+(1-x)\cdot log(1-y))
\end{equation}
\begin{equation}
J_{KL} = - \frac{1}{2} \cdot (1 + log\ \sigma^2 - \mu^2 - \sigma^2)
\end{equation}
\begin{equation}
J_{vae} = J_{xent} + J_{KL}
\end{equation}

\vspace{-0.5cm}
\subsubsection{Generator} After training, we use the trained decoder as a generator for batch generation of addresses. By sampling the 16-dimensional vector as a sample input in a standard normal distribution, the final generator outputs our ideal scan candidate. We set the sampling time $N$ to control the number of targets we expect to generate.

%
%

\subsection{Seed Classification}
Since IPv6 addresses include multiple addressing schemes, they are often intermixed in the seed set. Early classification of seeds with different structural patterns can help to improve the learning effect of the model on each structure of the address. The model then can generate addresses closer to the real structural pattern, which has greater possible activity. In this section, we will introduce two methods of seed classification that we used, including manual classification and unsupervised clustering.
\vspace{-0.2cm}
\subsubsection{Manual Classification}
In Section 3.1, we discussed the possible structural composition of the address. In this paper, we divide the address into four categories in Figure \ref{fig1}, including fixed IID, low 64-bit subnet, SLAAC EUI-64, and SLAAC Privacy. We perform feature matching on the active seed set to estimate the address category to which the seed belongs:

\begin{itemize} 

\item {\bfseries Fixed IID}. The last 16 nybbles have a unique consecutive 0 in the address. It is speculated that the last 16 nybbles may consist of the fixed IID.
\item {\bfseries Low 64-bit subnet}. The last 16 nybbles of the address have two or more consecutive 0 segments. It is speculated that it may consist of a subnet identifier and an IID.
\item {\bfseries SLAAC EUI-64}. The 23rd-26th nybbles of the address are {\bfseries fffe}.
\item {\bfseries SLAAC privacy}. After the statistics, the character appearance randomness of the last 16 nybbles is calculated, it is presumed to be a pseudo-random IID if the address has a high entropy value. We consider an address as SLAAC privacy if it has a greater entropy value than 0.8 (the highest is 1).
\end{itemize}

\subsubsection{Unsupervised Clustering}
We perform an entropy clustering method on the seed set, which was proposed by Gasser et al \cite{gasser2018clusters}. We applied the idea to the target generation algorithm for the first time.

In an address set S, we define the probability $P(x_i)$ for the character $x_i$ of the $i$-th nybble in an address, where $x\in\Omega=\{0,1,...,f\}$. Then by calculating the entropy value $H(X_i)$ for each nybble, we can get a fingerprint $F^a_b$ of the address set S:

\begin{equation}
F^a_b=(H(X_a),...,H(X_i),...,H(X_b))\\
\end{equation}
\begin{equation}
H(X_i) = -\frac{1}{4}\sum_{x\in \Omega}P(x_i)\cdot log\ P(x_i)
\end{equation}

where $a$ and $b$ are the first and the last considered nybble, respectively. Since /32 prefix is a large prefix that administrators usually use, which containing enough active addresses, we extract $F^9_{32}$ for each /32 prefix network address set (all addresses have the same first 8 nybbles in each network address set) and use the k-means algorithm to cluster each network fingerprint to find similar entropy fingerprint categories. 

\section{Evaluation}
In this section, we evaluate 6GCVAE effects. We will introduce the datasets used in the paper, the evaluation method, and our comparative experiment results.
\vspace{-0.3cm}

\subsection{Dataset}
Our experimental datasets are mainly from two parts, a daily updated public dataset IPv6 Hitlist and a measurement dataset CERN IPv6 2018. Table \ref{tab2} summarizes the datasets used in this paper. The public dataset IPv6 Hitlist is from the data scanning the IPv6 public list for daily active addresses, which is provided by Gasser et al \cite{gasser2018clusters}. In addition, we passively collected address sets under the China Education and Research Network from March to July 2018. We continued to scan and track the IPs that are still active until October 14, 2019 as our measurement dataset.

\vspace{-0.3cm}
\begin{table}
\setlength{\belowcaptionskip}{-0.1cm}
\caption{The detail of the two active address datasets we used in the paper.}\label{tab2}
\begin{center} 
\begin{tabular}{|c|c|c|c|}
\hline
Dataset & Seeds & Period & Collection Method\\
\hline
IPv6 Hitlist & 3,157,675 & October 14, 2019  & Public\\
CERN IPv6 2018 &  90,010 & March 2018 - July 2018 & Passive measurement\\
\hline
\end{tabular}
\end{center} 
\vspace{-1cm}
\end{table}

\vspace{-0.5cm}
\subsection{Evaluation Method}

\subsubsection{Scanning Method}To evaluate the activity of the generated address, we use the Zmapv6 tool  \cite{gasser2018clusters}  to perform ICMPv6, TCP/80, TCP/443, UDP/53, UDP/443 scans on the generated address. When the query sent by any scanning method gets a response, we will determine the address as active. Due to the difference in activity between hosts at different times, we maintain continuous scanning of the host for 3 days to ensure the accuracy of our method.
%
\vspace{-0.5cm}
\subsubsection{Evaluation Metric} Since IPv6 target generation is different from text generation tasks, we need to define a new evaluation metric for the address generative model. In the case of a given seed set, $N_{candidate}$ represents the number of the generated candidate set, $N_{hit}$ represents the number of generated active addresses, $N_{new}$ represents the generated address that is active and not in the seed set. Then the active hit rate $r_{hit}$ and active generation rate $r_{gen}$ of the model can be computed as 

\begin{equation}
r_{hit} = \frac{N_{hit}}{N_{candidate}} \times 100\% \qquad \qquad r_{gen} = \frac{N_{new}}{N_{candidate}}  \times 100\%
\end{equation}

We consider that $ r_ {hit} $ can represent the learning ability to learn from the seed set. $ r_ {gen} $ highlights the generation ability to generate new active addresses.

\subsection{Result of Seed Classification}
First, we summarize our seed classification. After manual classification, the seed will be classified into four categories. Table \ref{tab3} shows the classification details on the IPv6 Hitlist dataset. 

For the unsupervised clustering, we use the elbow method to find the number of clusters, $k$, plotting the sum of squared errors (SSE) for $k = \{1, ... , 20\}$. We selected the value $k=6$ for the point where increasing $k$ does not yield a relatively large reduction in SSE. Figure \ref{fig4} shows the results of the clustering.

\begin{table}
\vspace{-0.3cm}
\setlength{\belowcaptionskip}{-0.3cm}
\caption{The detail of manual classification on the IPv6 Hitlist dataset.}\label{tab3}
\begin{center} 
\begin{tabular}{|c|c|c|c|}
\hline
Category & Feature & Seeds & Percentage\\
\hline
Fixed IID & The last 16 nybbles have a consecutive 0 & 1,208,117  & 38.26\%\\
Low 64-bit Subnet & The last 16 nybbles have more consecutive 0 & 1,062,093 & 33.64\%\\
SLAAC EUI-64 &  The 23-26th nybbles  is fffe. & 279,458 & 8.85\%\\
SLAAC Privacy & Entropy value of the last 16 nybbles $> 0.8$& 608,007 & 19.25\%\\
\hline
Total &  - & 3,157,675 & 100\%\\
\hline
\end{tabular}
\end{center} 
\vspace{-1.9cm}
\end{table}

\begin{figure}
\setlength{\abovecaptionskip}{0.1cm}
\setlength{\belowcaptionskip}{-0.6cm}
\includegraphics[width=\textwidth]{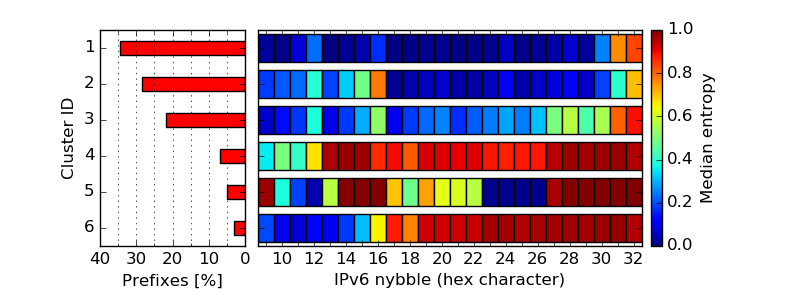}
\caption{The detail of unsupervised entropy clustering on the IPv6 Hitlist dataset. We obtained 6 clusters of all /32 prefix networks and the nybble distribution of each cluster.
} \label{fig4}
\end{figure}

It is worth noting that there is a certain relationship between the seed classification results. We know that the closer the $H(X_i)$ is to 0, the more likely the nybble is to be constant. The closer the $H(X_i)$ is to 1, the more random the nybble is. Therefore, in Figure \ref{fig4}, Cluster 1-3 may be fixed IID or a low 64-bit subnet addresses. Cluster 5 is likely to be SLAAC EUI-64 addresses with the {\bfseries fffe} flag. Cluster 4 and Cluster 6 are likely to be SLAAC privacy addresses because of the high entropy value of most of the nybbles.

After seed classification, we trained 6GCVAE with each category of seed sets. Table \ref{tab4} shows the effect of the model without seed classification, with manual classification and with unsupervised clustering on the IPv6 Hitlist dataset.

The model is trained by using the dataset or each category of seeds as a seed set and uses the generator to generate candidate targets after 1,000,000 samplings. We remove duplicate candidate targets and ultimately get a valid candidate set. The results show that seed classification can actually improve the performance of the model. Among them, the most generated addresses are manually classified Fixed IID and unsupervised clustered Cluster 2 respectively in the two methods. However, Low 64-bit subnet, SLAAC EUI-64, and Cluster 3-6 show a lower $r_{gen}$ due to the complex address structure or lack of training samples. In addition, the model has a characteristic on the generation of SLAAC privacy addresses. All generated hits are new active targets. Because of the high randomness of this kind of address, the model may learn a high random structure, resulting in the generated addresses which are without duplicates.

\begin{table}[t]
\caption{Model effect with 3 types of seed processing, including none of the seed classification, manual classification,  and unsupervised clustering.}\label{tab4}
\begin{center} 
\begin{tabular}{|c|c|c|c|c|c|c|}
\hline
Seed Classification & Category & $N_{candidate}$ & $N_{hit}$ &  $N_{new}$ & $r_{hit}$ & $r_{gen}$\\
\hline
None  & IPv6 Hitlist & 756,658 & 14,894 & 9,685 & 1.97\% & 1.28\%\\
\hline
           & Fixed IID & 412,181 & 32,589 & {\bfseries 17,933} & 7.91\% & {\bfseries 4.35\%}\\
Manual & Low 64-bit Subnet & 901,222& 7,092 & 5,450 & 0.79\% & 0.61\%\\
Classification & SLAAC EUI-64 & 981,204 & 1,299 & 1,263 & 0.13\% & 0.13\%\\
           & SLAAC Privacy & 999,920 & 13,351 & 13,351 & 1.34\% & 1.34\%\\
\hline
	& Cluster 1 & 526,542 & 25,235 & 12,364 & 4.79\% & 2.35\%\\
	& Cluster 2 & 450,919 & 57,245 & {\bfseries 35,508} & 12.70\% & {\bfseries 7.87\%}\\
Unsupervised & Cluster 3 & 759,617 & 5,273 & 2,404 & 0.69\% & 0.32\%\\
Clustering	& Cluster 4 & 985,390 & 6,605 & 6,309 & 0.67\% & 0.64\%\\
	& Cluster 5 & 832,917 & 1,748 & 845 & 0.21\% & 0.10\%\\
	& Cluster 6 & 968,178 & 1,193 & 994 & 0.12\% & 0.10\%\\
\hline
\end{tabular}
\end{center} 
\vspace{-0.3cm}
\end{table}

\vspace{-0.2cm}
\begin{table}[t]
\caption{The comparative experiments result by comparing with 5 conventional VAE models and Entropy/IP. Results show that unsupervised clustering reached the best performance in our experiments.}\label{tab5}
\begin{center} 
\begin{tabular}{|c|c|c|c|c|c|}
\hline
Model & $N_{candidate}$ & $N_{hit}$ &  $N_{new}$ & $r_{hit}$ & $r_{gen}$\\
\hline
FNN VAE & 1,000,000 & 68 & 68 & 0.007\% & 0.007\%\\
RNN VAE & 498,509 & 3,009 & 2,085 &  0.604\% & 0.418\%\\
Convolutional VAE & 595,475 & 4,432 & 2,856 & 0.744\% & 0.480\%\\
LSTM VAE & 478,660 & 4,464 & 3,203 & 0.933\% & 0.669\%\\
GRU VAE & 525,134 & 5,694  & 4,548 & 1.084\% & 0.866\%\\
\hline
Entropy/IP & 593,795 & 15,244 & 5,402 & 2.570\% & 0.910\%\\
6GCVAE & 756,658 & 14,894 & 9,685 & 1.970\% & 1.280\%\\
\hline
6GCVAE with Manual Classification & 557,653 & 28,957 & 15,870 & 5.193\% & 2.846\%\\
6GCVAE with Unsupervised Clustering & 571,330 & 54,915 & {\bfseries 31,376} & 9.611\% & {\bfseries 5.492\%}\\
\hline
\end{tabular}
\end{center} 
\vspace{-0.8cm}
\end{table}

\subsection{Comparing with Conventional VAE Models}
In order to verify the superiority of 6GCVAE, we built the baseline of the conventional VAE models by replacing the key components gated convolutional layer of 6GCVAE and compared them with our model. We also use the generator for 1,000,000 samples after training the model with the IPv6 Hitlist dataset. Table \ref{tab5} summarizes the results of the comparative experiments. The results show that due to the inability of feedforward neural networks to well capture semantic information, the FNN VAE displays a difficulty to complete the IPv6 target generation task. RNN VAE and Convolutional VAE only focus on sequence relationships or structure information, thus causing lower hits. By promoting the simple RNN layer to LSTM or GRU, the VAE model gets better performance than RNN VAE. Finally, 6GCVAE performs best under this task because of learning both the key segment structure and segment relationship information of an address.

%

\begin{figure}[t]
\vspace{-0.2cm}
\centering
\subfigure[IPv6 Hitlist]{       
\label{1}
\begin{minipage}[t]{0.47\linewidth}
\centering
\includegraphics[width=5.9cm]{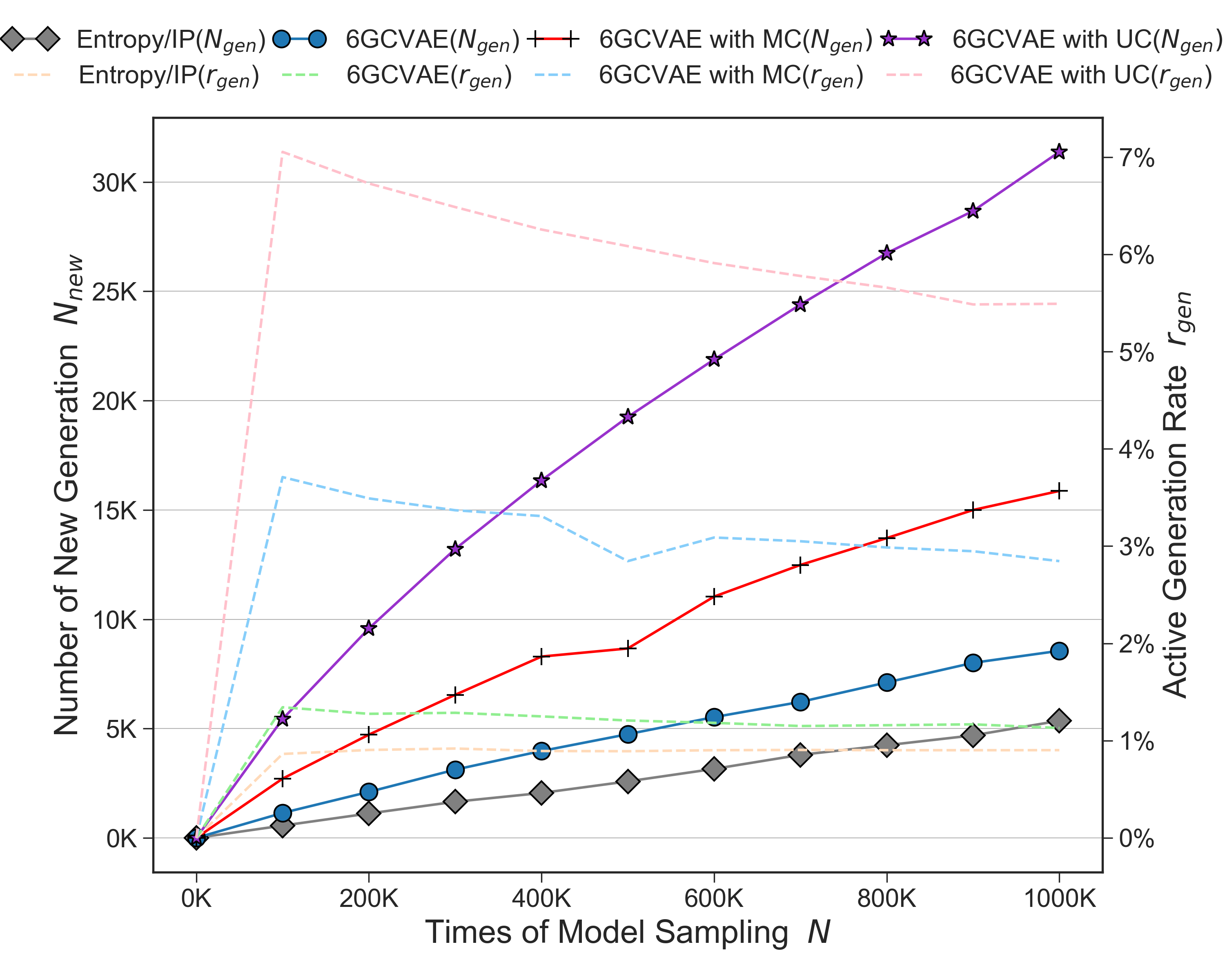}
\end{minipage}
}
\subfigure[CERN IPv6 2018]{ 
\label{2}
\begin{minipage}[t]{0.47\linewidth}
\centering

\includegraphics[width=5.9cm]{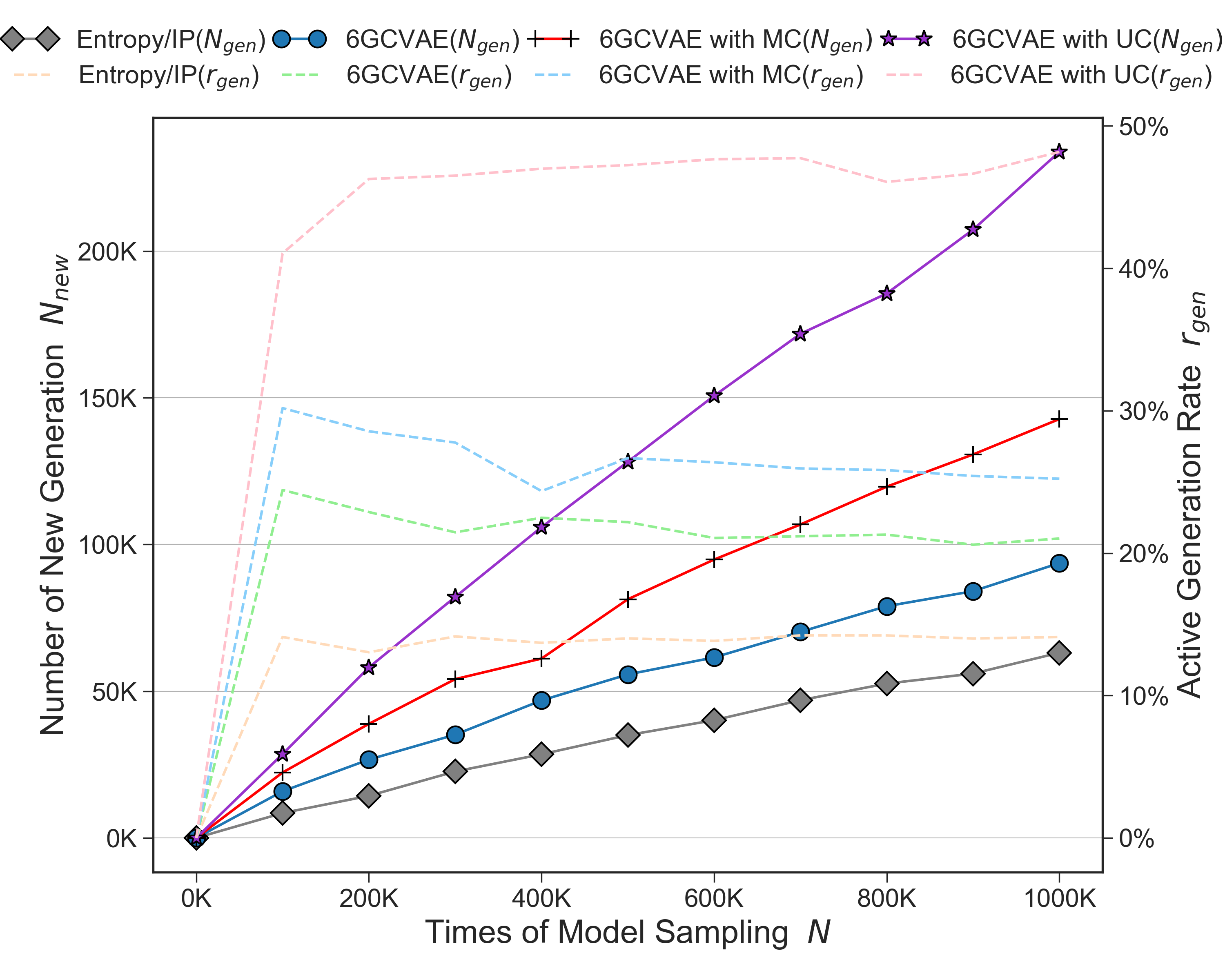}
\end{minipage}
}
\centering
\setlength{\abovecaptionskip}{-0.1cm}
\caption{The comparative experiments result by comparing with Entropy/IP on the two datasets. $N_{new}$ and $r_{gen}$ are evaluated under the different sampling times $N$.}
\label{fig5}
\vspace{-0.5cm}
\end{figure}

\subsection{Comparing with Entropy/IP}
\vspace{-0.1cm}
Entropy/IP \cite{foremski2016entropy} is the current state of the art address generation tool that can also efficiently generate active IPv6 targets. We compare the effects of 6GCVAE with Entropy/IP by training model and sampling 1,000,000 times for target generation as usual. As shown in Table \ref{tab5}, the experimental results show that our model outperformed Entropy/IP under the IPv6 Hitlist dataset. Although the $r_{hit}$ of Entropy/IP is higher, its lower $r_{gen}$ indicates that it generates more addresses that are duplicated in the dataset.

For representing the final effect of each seed classification method, we control the generation ratio of each type of address through $r_{gen}$ to maximum the generation of new active target $N_{new}$. The ratio can be represented as $(r_{gen_1}:r_{gen_2}:...:r_{gen_i})$, where $i$ represents the category id of a seed classification method. Finally, we set the total sampling number $N$ in each round of experiments and control the generation number of each category of seed set through the ratio. We then reached the best experimental results in Table \ref{tab5}. 6GCVAE has been greatly improved with seed classification. 

In addition, in Figure \ref{fig5}, we evaluated $N_{new}$ and $r_{gen}$ by changing the sampling times $N$ on the two datasets, which can prove the general generation ability of the models. Results indicate that our approach reaches a better performance than Entropy/IP. 6GCVAE found 1.60-1.79 times more hits than Entropy/IP. Under manual classification (MC) and unsupervised clustering (UC), the $N_{new}$ of 6GCVAE has been improved 1.52-1.85 and 2.50-3.67 times respectively. The seed classification methods have a higher $r_{gen}$ than all other approaches. Unsupervised clustering reached the best performance in our experiments. 

\vspace{-0.4cm}
\section{Conclusion}
\vspace{-0.3cm}
In this paper, we explored the challenges of IPv6 target generation tasks. Our work achieved a new model 6GCVAE by constructing a gated convolutional Variational Autoencoder. In addition, we introduce two kinds of seed classification techniques, which effectively improve the address generation performance of the deep learning model. The results show that 6GCVAE is superior to the previous conventional VAE models. The address generation quality of 6GCVAE is better than the state of the art target generation algorithm Entropy/IP.

\vspace{-0.3cm}
\subsubsection{Acknowledgements}
This work is supported by The National Key Research and Development Program of China (No.2016QY05X1000) and The National Natural Science Foundation of China (No. U1636217) and Key research and Development Program for Guangdong Province under grant No. 2019B010137003.
\vspace{-0.6cm}

%
%
%
 \bibliographystyle{splncs04}
 \bibliography{mybibliography}
%




\end{document}